# EEG and ECG changes during deep-sea manned submersible operation

Haifei Yang[1], Lu Shi[1, 2*], Feng Liu[3], Yanmeng Zhang[2], Baohua Liu[4], Yangyang Li[2], Zhongyuan Shi[2] and Shuyao Zhou[2]

[1]SJTU-CU International Cooperative Research Center, School of Naval Architecture, Ocean and Civil Engineering, Shanghai Jiao Tong University, No.800 Dong Chuan Road, Shanghai 200240, China

[2]Institute of Underwater Technology, Shanghai Jiao Tong University, No.1500 Long Wu Road, Shanghai 200231, China

[3]China Ocean Mineral Resources R&D Association (COMRA), No.1 Fu Xing Men Wai Road, Beijing 10083, China

[4]China National Deep Sea Center (NDSC), No.1 Wei Yang Road, Qingdao 266237, China

*Correspondence: shilu@sjtu.edu.cn

## Abstract

**Background:** Deep-sea manned submersible operation could induce mental workload and influence neurophysiological measures. Psychophysiological responses to submersible operation are not well known. The main aim of this study was to investigate changes in EEG and ECG components and subjective mental stress of pilots during submersible operation.

**Methods:** There were 6 experienced submersible pilots who performed a 3 h submersible operation task composed of 5 subtasks. Electroencephalogram (EEG) and electrocardiogram (ECG) was recorded before the operation task, after 1.5 h and 2.5 h operation, and after the task. Subjective ratings of mental stress were also conducted at these time points.

**Results:** HR and scores on subjective stressed scale increased during the task compared to baseline ($P<0.05$). LF/HF ratio at 1.5 h were higher than those at Baseline ($P<0.05$) and 2.5 h ($P<0.05$). Relative theta power at the Cz site increased ($P<0.01$) and relative alpha power decreased ($P<0.01$) at 2.5 h compared to values at Baseline. Alpha attenuation coefficient (AAC, ratio of mean alpha power during eyes closed versus eyes open) at 2.5 h and after the task were lower compared to baseline and 1.5 h ($P<0.05$ or less).

**Conclusions:** Submersible operation resulted in an increased HR in association with mental stress, alterations in autonomic activity and EEG changes that expressed variations in mental workload. Brain arousal level declined during the later operation period.

## 1. Introduction

Deep-sea manned submersible operation is a complex activity for submersible pilots. During the submersible operation, apart from the primary tasks (dive, navigation, communication), submersible pilots have to plan for activities, supervise the system status, and anticipate future tasks. These requirements for submersible pilots and those for aircraft pilots have much in common [33]. Thus, analogously, submersible operation requires a high level of cognitive activity associated with various stress factors, including time constraints, safety threats, and environmental factors [27]. These constraints induce a significant mental workload, which "is the general term used to describe the mental cost of accomplishing task requirements"

[12].

Mental workload has been examined with various methods in recent decades, such as performance evaluation, subjective reporting, and psychophysiological measurements. In a complex task environment, performance measures often cannot index workload, particularly for each subtask [33]. As a mental workload is a subjective state as well, subjective reporting generally provides a good indication of total workload. Nevertheless, there remain some methodological issues with subjective rating, such as trading off the intrusiveness of online or live ratings against the retrospective bias of post-task ratings [36]. Psychophysiological measures, for instance, electroencephalogram (EEG), electrocardiogram (ECG), and eye activity, are found to be robust candidates for operators workload evaluation [10,34].

Numerous studies showed variations in the power spectrum of EEG bands during mental workload. There are much findings indicate that increased workload leads to increased frontal theta (4-8 Hz) activity, for example, in traffic control tasks [2], working memory tasks [7,8] or flight simulation tasks [28]. Alpha (8–13 Hz) band activity decreases in frontal and central areas [2,7,30] or parietal area [34] when cognitive demand intensifies or when the task becomes more difficult.

From a cardiovascular point of view, like piloting aircrafts, operating submersible requires autonomic adaptation, probably giving rise to alterations in heart rate (HR), respiration, and blood pressure [30]. Heart rate variability (HRV), indicating autonomic nervous system activity, has been found to respond reliably to changes in workload and mental effort in some studies [27,30,34].

Submersible pilots managing complex work environments are exposed to different stressors, like confined space and time pressure, which would induce mental stress. Mental stress can elicit prompt and consistent increases in arterial blood pressure and heart rate [4]. Hjortskov et al. suggested HRV is a more sensitive and selective measure of mental stress than blood pressure [14].

Furthermore, according to Borghini et al., during piloting or driving, the sequence of internal brain states can be described as a series of transitions from the "alert state" to the "fatigued" one, and from there to the "drowsy state" [1]. Thus the detection of brain state is important during the submersible operation. Alpha Attenuation Test (AAT) is able to quantify arousal level [29], which could explain the fatigue effect to some extent [21].

However, there are few studies concerning submersible pilots because the low number of manned submersibles and its pilots. Here we studied EEG and ECG changes in the pilots of Jiaolong, a Chinese manned deep-sea research submersible that can dive to and operate at a depth of over 7,000 m. Jiaolong was put into service in 2010 and reached a depth of 7,062 meters (23,169 feet) in the Mariana Trench in the western Pacific Ocean in 2012. Operating Jiaolong, especially in a deep dive, have differences in environment and task types compared with aircraft piloting. For instance, in an actual undersea dive, as subjected to vibration, noise and sometimes rocking motion, pilots usually have to concentrate on piloting in dark environment for about 8 to 10 hours. During cruise, they ought to watch out for hydrothermal vent and obstacle through portholes by sight. Robot hands manipulation is another specific task which requires high visual attention. These differences and the complicated mechanism among human, machine and environment interaction determine that we cannot take results from aircraft piloting or other operation types for granted. This study expects to explore the psychologic and physiologic changes in pilot and analyze these responses to predict mental state in case of mental overload and fatigue.

We hypothesized that the workload induced by the complex and stressful submersible operation task would modify EEG parameters and cardiovascular activity and induce fatigue during the task. So the aims of the present study were to: 1) document the EEG and ECG, evaluate their changes by time and(or) frequency domain analysis, in submersible pilots during a 3 h submersible operation task; 2) evaluate mental stress and fatigue during this task; 3) assess correlations between ECG components, EEG components and mental stress.

## 2. Methods

**Subjects**

Ethical approval for this study was provided by the ethical committee of Chinese Underwater Technology Institute of Shanghai Jiao Tong University. All pilots read and signed a consent form before the experiment. Six healthy male submersible pilots (age: 27.6±1.97 yr, height: 170.8±5.61 cm, body weight: 73.8±7.64 kg) participated in this study. They all meet the criteria for selection of deep-sea manned submersible pilots and had all been trained to pilot Jiaolong submersible for over 2 years. In order to control for confounding factors, pilots were asked to refrain from drinking coffee or tea, eating, and smoking for at least 2 h before the operation task.

**Task environment**

The submersible operation task in this study was conducted in an experimental pool at China Ship Scientific Research Center (Jiangsu, China). This pool was used for conducting diving test and submersible pilots' daily training.

The cabin of Jiaolong was capable of carrying two pilots and one observer. The volume of this cabin was 4.8 $m^3$. During the entire task, the cabin temperature was 15-18 ℃, humidity 65-70 %, oxygen concentration 21 % and carbon dioxide concentration less than 0.03 %.

**Physiological measurement**

The EEG and ECG was conducted simultaneously and lasted 4 min each time. The recording protocol was designed to make sure at least one pilot would execute the corresponding subtask during operation and thus ensure safety (**Fig.1**). In the first/last 2 minute of recordings for pilot A/B, he was asked to repeatedly close and open his eyes every 30 sec. This protocol was based on AAT developed by Michimori et al [29] to access sleepiness. In the last/first 2 minute of recordings for pilot A/B, he continued to execute the corresponding subtask, or just sat quietly during baseline and recovery recordings.

EEG was recorded at Cz in the 10-20 system by a BIOPAC ® MP150 (BIOPAC Systems Inc., U.S.). ECG was recorded using a BIOPAC ® MP30 (BIOPAC Systems Inc., U.S.). A modified lead II configuration was employed for ECG recording; the negative lead was placed on the central forehead, ground just under the supraclavicular fossa (between the two collarbones), and positive lead on the left side over the lower rib. All signals were sampled at 500 Hz and collected on the experimenter's computer using Ag/AgCl disposable electrodes.

**Subjective ratings**

Subjective ratings were all conducted immediately after the EEG and ECG recordings. Visual analogue scale (VAS) was employed for subjective ratings of mental stress of pilots. VAS is a psychometric response scale which can be used in questionnaires [13]. The following six 11-point scales (0=not at all,

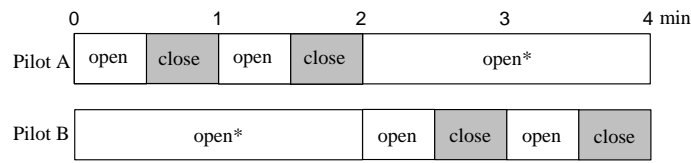

**Fig.1** EEG and ECG recording protocol. Note: open: opening eyes; close: closing eyes; *: sitting quietly during baseline and recovery recording/executing the corresponding subtask during operation.

10=extremely) were used: stressed, tense, exhausted, concentrated, stimulated and happy [14,26].

**Procedure**

The experiment last for 3 d and we have two pilots to be engaged in the operation task each day. On the day before the first day of experiment, all pilots were well informed of the procedure of the operation task and necessary descriptions mainly about ECG and EEG measurements. The operation task was carried out in the afternoon only, to control for circadian effects.

The following steps were performed in each day of the three-day experiment. The baseline EEG, ECG and subjective ratings of the two pilots were assessed 30 min before the task in a quiet room. Then the two pilots were escorted to Jiaolong and seated in the cabin. After all conditions for the submersible dive were checked, the submersible was hung up and put into the pool. Then our two pilots started to pilot the submersible. The two pilots were instructed to pilot and manipulate the submersible to complete 5 subtasks successively: diving, touching bottom, cruise, collecting samples of soil using robot hands and coming up to the surface. The entire operation task took 3 h. These two pilots were required to conduct EEG and ECG recordings and then subjective ratings after 1.5 h and 2.5 h operation. The former (1.5 h) was during the cruise subtask and the latter (2.5 h) was during the robot hands manipulation subtask. Subjective ratings were carried out alternately, and then pilots continued to execute the subtasks. During the operation task, experimenters used underwater communication system to communicate with the pilots. After the pilots completed the whole task and come out of the cabin, they were escorted to a quiet room to rest for 10 min and then EEG and ECG recording and subjective ratings were administered.

**Data processing and analysis**

A 0.5-35Hz band-pass filter was applied on the EEG signals. Then, Data were segmented into 4.096 s (2048 samples) epochs with 50% overlap. EOG artifact correction algorithms [19,20] were then applied on each epoch data, and semi-automatic artefact detection was applied on the remaining amplitude (range 100 mV) on Acknowledge 4.2 (BIOPAC Systems Inc., U.S.). The data were further analyzed using fast Fourier transform (FFT, hanning window) for three standard bandwidths: theta (4-8 Hz), alpha (8-13 Hz), and beta (13-30 Hz). Relative power spectrum density (PSD) of these three frequency bands (percentages of the total power of 4-30 Hz) derived from the last/first 2 min for pilot A/B was calculated. Alpha attenuation coefficient (AAC) was calculated as the ratio of mean alpha power during eyes closed versus eyes open [29].

The ECG signals were analyzed using Kubios HRV (University of Eastern Finland, Finland). The software automatically removed baseline trend and detected R peaks from the ECG data. In time domain analysis, the mean HR, the standard deviation (SD) between normal-to-normal intervals (SDNN), and the root mean

**Table 1.** Means (SD) for scores on subjective experiences of stress before the operation task (Baseline), after 1.5 h operation (1.5 h), after 2.5 h operation (2.5 h) and after the operation task (Recovery).

| variables and conditions | Baseline | 1.5 h | 2.5 h | Recovery |
|---|---|---|---|---|
| stressed | 0.5 (0.7) | 1.3 (1.0)* | 1.4 (1.1)* | 0.7 (1.1) |
| tense | 0.3 (0.3) | 0.7 (0.8) | 0.8 (1.1) | 0.7 (1.1) |
| exhausted | 0.2 (0.3) | 0.7 (0.9) | 0.6 (1.0) | 0.7 (1.0) |
| concentrated | 8.0 (1.5) | 7.6 (1.3) | 7.5 (1.5) | 7.6 (1.2) |
| stimulated | 0.4 (0.5) | 0.9 (0.8) | 0.7 (1.0) | 0.7 (0.8) |
| happy | 6.8 (1.8) | 6.4 (2.6) | 6.9 (2.5) | 6.6 (2.2) |

* values recorded during operation vs. Baseline (*$P<0.05$)

square of successive differences of successive normal-to-normal intervals (RMSSD) were calculated. In frequency domain analysis, the following parameters were calculated: power in the high frequency (LF) range (from 0.04 to 0.15 Hz), power in the high frequency (HF) range (from 0.15 to 0.4 Hz), and the LF/HF ratio. HF is associated with mechanical and reflex respiratory activity components and is usually interpreted as reflecting oscillations caused mainly by changes in vagal tone. LF reflects both sympathetic and vagal activity. The LF/HF ratio is considered a reflection of the sympatho-vagal balance or sympathetic modulation [31].

*Statistical analysis:* Data of four time points, including before the operation task (Baseline), after 1.5 h operation (1.5 h), after 2.5 h operation (2.5 h) and after the operation task (Recovery), were analyzed. After normality assessment by Shapiro-Wilk test, differences between these values for all psychophysiological indices were assessed by analysis of variance (ANOVA) with repeated measures with Bonferroni multiple comparisons, or Friedmann test with Dunn's multiple comparison for non-parametric data. Linear regression was used to assess quantitative relationships between EEG, ECG components and scores in mental stress. Statistical tests were assessed using IBM® SPSS statistics 20 for Windows® software. The level of statistical significance was set at $P=0.05$.

## 3. Results

Compared to Baseline, HR at 1.5 h and 2.5 h significantly increased ($P<0.05$) and then deceased back to baseline level at Recovery. We did not observe a time effect for LF+HF. LF power and LF/HF ratio at 1.5 h were significantly higher than those at Baseline ($P<0.05$) and 2.5 h ($P<0.05$). Inversely, HF power significantly decreased at 1.5 h ($P<0.05$) compared to Baseline and then significantly increased at 2.5 h ($P<0.05$) compared to 1.5 h. No significant difference was seen for SDNN and RMSSD between values at Baseline and other points (**Fig.2**)

Relative theta power at the Cz site increased during the operation at 2.5 h compared to the value at Baseline ($P<0.01$). For relative alpha power at the Cz site, the 2.5 h value are higher than Baseline value ($P<0.01$). No significant difference was observed for relative power in Beta band between Baseline and other points.

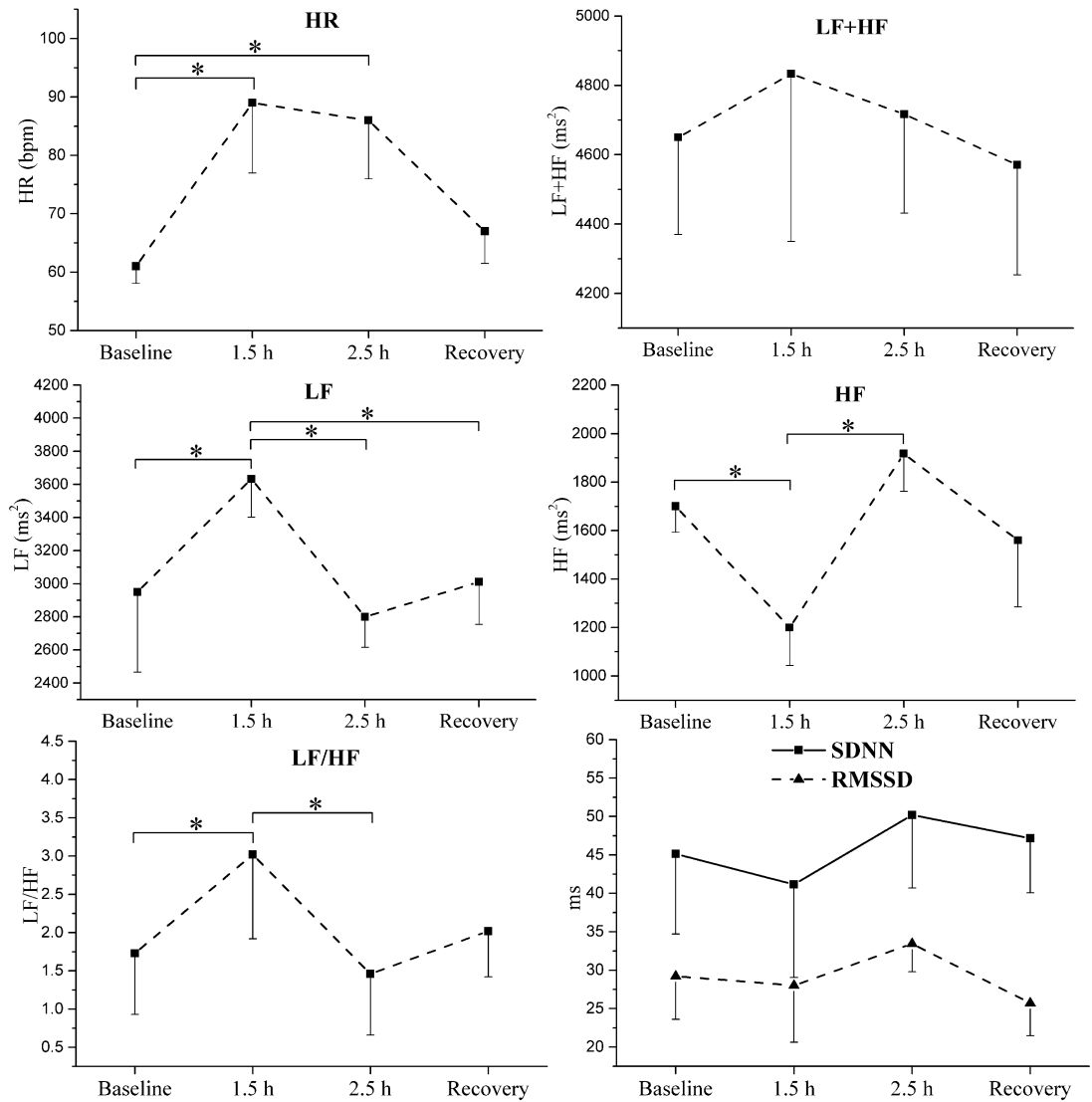

**Fig.2** Heart rate and heart rate variability components in temporal and frequency domains before the operation task (Baseline), after 1.5 h operation (1.5 h), after 2.5 h operation (2.5 h) and after the operation task (Recovery). (* $P$ <0.05).

For AAC, the ratio of mean alpha power during eyes closed versus eyes open, the 2.5 h and Recovery values were significantly lower than those at Baseline and 1.5 h ($P$<0.05 or less). (**Fig.3**)

**Table 1** includes the results of subjective ratings. Mean score on subjective stressed scale at 1.5 h and 2.5 h were both significantly increased compared to Baseline ($P$<0.05). No differences were seen between different time points for the other stress scales, i.e. tensed, exhausted, concentrated, stimulated and happy. The only observed significant linear regression was between HR and scores on stressed scale ($P$=0.01).

## 4. Discussion

The main aim of the present study was to investigate the changes in EEG and ECG components during submersible operation. These data suggest some significant changes of EEG and ECG components at 1.5 h,

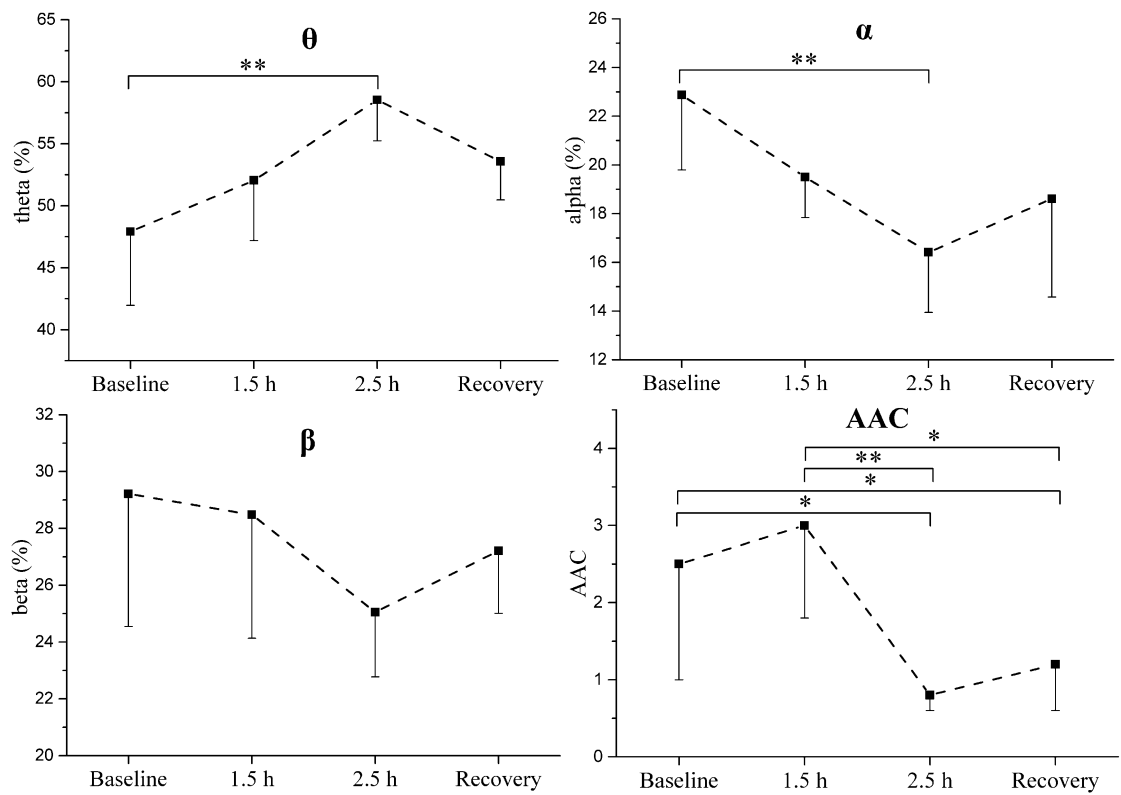

**Fig.3** Relative power in theta(θ), alpha(α), beta(β) band and alpha attenuation coefficient (AAC) before the operation task (Baseline), after 1.5 h operation (1.5 h), after 2.5 h operation (2.5 h) and after the operation task (Recovery) . (* $P <0.05$,** $P <0.01$).

2.5 h and Recovery points. These changes were discussed in the following section.

**HR and HRV**

The augmentation on HR during the operation task could be related to increased cognitive load, emotional stress and temporal pressure. The relation with emotional stress could be confirmed by the significant linear regression between HR and scores on stressed scale. The results of HR data reinforce previous work indicating that HR can be sensitive to workload [6,27,30,34].

The increased LF power, LF/HF ratio and decreased HF power at 1.5 h suggested an increase of sympathetic activity and a vagal withdrawal, which was consistent with results of previous study suggesting that increased LF/HF ratio indicated mental stress [14] or task load [6]. However, although scores on stressed scale and task load were still higher than baseline at 2.5 h, LF/HF ratio decreased to baseline at this point. This was probably because as pilots were gradually adapted to the task environment and familiar with the operation sequences, time pressure relieved and cardiac autonomic activity produced a new sympatho-vagal balance.

The results of HRV analysis suggest HRV components might not be reliable enough to indicate mental workload increasing during submersible piloting in this study. Nickel and Nachreiner proposed that HRV is an indicator for time pressure or emotional strain, not for mental workload [24]. According to the study of a ship navigator by Murai et al. [23], the LF/HF ratio level and response correlated strongly with emotional factors and varied in different cases. Our further studies need to consider more factors, like emotional strain, pilots' proficiency and task sequences to explore their potential influences on autonomic activity while using HRV to indicate mental workload.

**EEG**

EEG power reflects the amount of neurons that discharge at the same time [18]. The discharge generates oscillatory activities that occur more frequently during more than less demanding tasks [16]; thus they are thought to be related to the cortical resources employed for information processing [17]. Here, we observed a rising trend of theta band power during submersible operation and it reduced to baseline level at Recovery. The theta band activity was reported to be associated with working memory load [8], mental processing [9], mental calculations during flight [10] and complex cognitive activity [2,28]. While piloting the submersible, pilots needed to continuously monitor the system and environment, which required much visual attention. This corresponded to previous studies describing modifications of the theta band power may reflect the effort at visual attention to the flight instruments during instrument flight [5,10]. The theta band activity recorded by the Cz electrode at 2.5 h was significantly greater than that recorded at Baseline, suggesting that robot hands manipulation required the most visual attention and cognitive activity.

Power in the alpha band at the Cz site was decreasing throughout the operation task compared to the baseline value, which corresponded to the inverse relationships between alpha power at Cz and task difficulty reported by earlier studies [2,7,30] .. Alpha has been linked to different type of idling [25], cortical inhibition [32], and default mode brain activity [15]. These would all roughly be consistent with alpha power varying with different levels of workload [3]. However, in aircraft piloting, alpha band activity was reported to lack sensitivity in response to requirements of piloting during different flight segments [10]. The greatest variation in alpha band activity and the theta band activity discussed above at 2.5 h indicate that the mental workload at 2.5 h seemed higher than that at 1.5 h, which could not derived from the ECG components. EEG seems to be a more sensitive indicator of mental workload than ECG in our operation task.

The AAC at 2.5 h was lower than that at Baseline and 1.5 h, and it didn't increase back to baseline level during Recovery period. These results indicate a decline in brain arousal level [29] after 2.5 h submersible operation, and it continue to exist after the entire task. The decreased AAC suggest a low level of fatigue might appeared after 2.5 h submersible operation, although the low scores on exhausted scale (mean value<1, **Table 1**) indicating very slight subjective fatigue. However, mental fatigue at lower levels could also slow sensorimotor functions and impair information processing, diminishing pilots' capability to pilot the submersible [22]. Thus, a reasonable submersible operation scenario should include enough regular rest time, especially during a long time operation task.

**Limitations and future studies**

This study was designed as a preliminary investigation since the few relevant previous studies and the specificity of our task environment. Because of our small *n*, some of our statistical findings may have resulted from response peculiarities of particular participants. However, the small *n* is reasonable due to the specific small number of Jiaolong pilots. Also, EEG signals of more electrodes on the scalp should be recorded and analyzed in the future studies. Neurophysiological methods give us more information that cannot be obtained from the subjective methods [35]. In the future research, we aim to mine more data to establish a specific framework based on neurophysiological indices to measure submersible pilots' workload and thus predict their performance.

## 5. Conclusions

The present study explored the changes of EEG and ECG components and subjective mental stress during a submersible operation task. Submersible operation involved an increased HR in association with subjective experience of stress. The operation task also resulted in the alterations in automatic nervous activity, while increased theta band activity and decreased alpha band activity indicated variations in mental workload. During the operation, a decline in brain arousal level in the later operation period might give a sign of mental fatigue. Nevertheless, supplementary studies with higher time precision and more detailed data are necessary to assess the mental workload and fatigue during submersible operation. These results might give better references in monitoring pilots' mental state, planning task schedule and optimizing human machine interface of the submersible.

## Acknowledgement

The study was supported in part by Natural Science Foundation of Shanghai (12ZR1415300), International Science & Technology Cooperation Program of China (2012DFA20970) and Public science and technology research funds projects of ocean (201205035-11). We would like to thank the personnel of China Ship Scientific Research Center (Jiangsu, China) and National Deep Sea Center (Shandong, China) for their technical assistance.